\documentclass[a4paper,11pt]{article}
\usepackage{pos}
\usepackage{amsmath}
\usepackage{tikz}

\newcommand{\rf}[1]{(\ref{#1})}
\newcommand{\rff}[1]{\figurename~\ref{#1}}
\newcommand{\rfs}[1]{Sec.~\ref{#1}}

\newcommand{\rb}[1]{\left(#1\right)}

\newcommand{\prt}{\partial}

\newcommand{\bigo}{\mathcal{O}}
\newcommand{\amp}{\mathcal{A}}
\newcommand{\ctns}{\mathcal{K}}
\newcommand{\cder}{\mathcal{D}}
\newcommand{\cu}{\mathrm{i}}

\newcommand{\anb}[1]{\langle#1\rangle}
\newcommand{\sqb}[1]{[#1]}
\newcommand{\anr}[1]{|#1\rangle}
\newcommand{\sqr}[1]{|#1]}

\newcommand{\sql}[1]{[#1|}

\title{Scalar-Vector Effective Field Theories from Soft Limits}
\ShortTitle{Scalar-Vector EFTs from Soft Limits}

\author*[a,\dagger]{Filip Přeučil}

\affiliation[a]{Institute of Particle and Nuclear Physics, Charles University,\\
V Holešovičkách 2, 180 00 Prague 8, Czech Republic}

\note[$\dagger$]{In collaboration with Karol Kampf, Jiří Novotný and Jaroslav Trnka}

\emailAdd{preucil@ipnp.mff.cuni.cz}

\abstract{We give an overview of the implementation of the soft-bootstrap method applied to the landscape of theories where the Special Galileon couples to a massless vector particle. We also describe the corresponding traditional Lagrangian approach for this model, which takes into account the formal geometrical interpretation of the Special Galileon as fluctuations of a $D$-dimensional brane embedded in a $2D$-dimensional flat space.}

\FullConference{%
  40th International Conference on High Energy physics - ICHEP2020\\
  July 28 - August 6, 2020\\
  Prague, Czech Republic (virtual meeting)
}

\begin{document}
\maketitle
\section{Introduction}
In the last few decades, the on-shell amplitude methods have undergone a considerable development, and many new and surprising results have been obtained (for a recent review and a complete list of references, see \cite{Elvang:2015rqa}). These developments concern not only the originally considered (SUSY) gauge theories and gravity, but also the class of non-renormalizable effective field theories (EFTs). Lagrangians of EFTs are organized as infinite towers of vertices with increasing mass dimension, and their form is fixed by the relevant symmetries. In some cases, the symmetry requirements are so strong that the Lagrangian of the corresponding EFT is unique up to a finite number of free coupling constants. This is typically the case of spontaneously broken symmetries, when the effective theory describes the dynamics of the Goldstone bosons. In such a case, the $S$-matrix of the theory has a peculiar IR behavior, which is expressed in terms of soft theorems for the scattering amplitudes. It often appears that these soft theorems can be taken as an alternative on-shell definition of the theory itself \cite{Cheung:2014dqa} and can be used for the recursive reconstruction of the tree-level $S$-matrix \cite{Cheung:2015ota}. Therefore, the on-shell amplitude methods in conjunction with soft theorems provide us with a powerful tool for exploration of the landscape of EFTs and for finding new EFTs with interesting properties. This program, dubbed as the soft bootstrap, has been completed for the single flavor scalar EFTs in \cite{Cheung:2016drk}, where the the classification of the exceptional theories with enhanced soft limits has been performed. The case of SUSY EFTs has been explored by these methods in \cite{Elvang:2018dco}, and the systematic studies of multi-flavor scalar theories have been initiated in \cite{Kampf:2020tne}.

In this contribution, we present our preliminary results illustrating the power and limitations of the method in the case of the coupling a massless vector particle to a massless scalar with Galileon power counting\footnote{This hypothetical theory was first mentioned in \cite{Elvang:2018dco,Bonifacio:2019rpv}.}. Physically, such an EFT might correspond to the coupling of the photon to the modified gravity in the decoupling limit, where the only interacting degree of freedom is the Galileon, or to the interaction of the vector and scalar degrees of freedom of the massive gravity near the decoupling limit.

\section{Bootstrap method}\label{s:bstr}
In this section, we describe briefly the application of the bootstrap method to the exploration of the landscape of EFTs containing a massless scalar (sGal) and a massless vector (we call it the BI photon in what follows). The sGal nature of the scalar means that tree amplitudes $\amp$ should posses an enhanced single scalar soft limit, i.e. $\amp=\bigo\rb{p^3}$, where $p\to 0$ is the momentum of any scalar leg. Similarly, the presence of the BI photon assumes for the amplitudes the properties similar to the pure BI theory, namely the helicity conservation and some sort of a multichiral soft limit \cite{Cheung:2018oki}.

\subsection{Method description}
The idea of the bootstrap method is to start with some lowest-order seed amplitudes (which are contact by construction), glue them together to ensure the right factorization, and then add some contact terms with free constants. These are to be determined using some additional information, in our case a certain set of soft limits. The whole process is then iterated as far as we can computatively go to construct higher-order amplitudes. For an illustration of the first iteration, see \rff{f:glue}.
\begin{figure}[ht]
\begin{center}
\begin{tikzpicture}[x=25pt, y=25pt]
\pgfmathsetmacro{\sc}{2/sqrt(3)};

\def\spc{.75};
\node at (0,0) {$\amp$};
\node at (\spc,0) {$=$};
\def\c1{2*\spc+1};
\draw (\c1-1,0) -- (\c1+1,0);
\draw (\c1,-1) -- (\c1,1);

\node at (3*\spc+2,0) {$\otimes$};
\def\c2{4*\spc+3};
\draw (\c2-1,0) -- (\c2+1,0);
\draw (\c2,-1) -- (\c2,1);

\node at (5*\spc+4,0) {$+$};
\def\c3{6*\spc+4};
\draw (\c3,0) -- (\c3+2*\sc,0);
\draw (\c3+.5*\sc,-1) -- (\c3+1.5*\sc,1);
\draw (\c3+1.5*\sc,-1) -- (\c3+.5*\sc,1);
\end{tikzpicture}
\end{center}
\caption{First iteration of the bootstrap method, construction of a 6-point amplitude from seed ones.}
\label{f:glue}
\end{figure}

\subsection{Power counting parameter}
Let us consider a tree amplitude $\amp$ with mass dimension $d$ and $n$ external legs, composed of vertices $V_i$ with mass dimensions $d_i$ and with $n_i$ external legs. Then it holds
\begin{equation}
d-2=\sum_i(d_i-2),
\hspace{90.157pt}
n-2=\sum_i(n_i-2).
\label{e:pcp}
\end{equation}
We define the power counting parameters $\varrho_\amp$ and $\varrho_i$ for the amplitude $\amp$ and for the vertices $V_i$ as
\begin{equation}
\varrho_\amp\equiv\frac{d-2}{n-2},
\hspace{105.313pt}
\varrho_i\equiv\frac{d_i-2}{n_i-2}.
\end{equation}
It is self-evident that all the tree amplitudes of an EFT with vertices that all have the same parameter $\varrho_i=\varrho$ also have $\varrho_\amp=\varrho$. We call such EFT single-$\varrho$ theory in what follows\footnote{Among single-$\varrho$ EFTs, there are distinguished theories such as the NLSM, DBI, BI, or Galileons.}. The question we are trying to answer is whether there is a unique theory with the Galileon-like power counting $\varrho = 2$ containing the sGal coupled to the BI photon.

\subsection{Contact terms and seed amplitudes}
Since we consider only the massless theories, we can employ the massless spinor-helicity formalism. The principles of Lorentz invariance and locality imply that contact amplitudes are polynomials in square and angle spinor brackets, and they are constrained by the power counting and the little group scaling. In a single-$\varrho$ theory, the mass dimension $d$ of any contact $n$-point amplitude $\amp$ should be $d=\varrho(n-2)+2$. Also, the amplitude should scale as $\amp\to z^{2h_i}\amp$ whenever the $i$-th leg (with helicity $h_i$) is scaled using $\sqr{i}\to z\sqr{i}$ and $\anr{i}\to z^{-1}\anr{i}$.

Let us now classify the seed amplitudes. Assuming the helicity conservation\footnote{This means that the number of helicity-plus and helicity-minus BI photons is the same when we assume all the particles as outgoing.} and the parity conservation, there are three 4-point seed amplitudes possible\footnote{Note that these amplitudes obey the sGal limit and also the multichiral soft limits in the sense of ref. \cite{Cheung:2018oki}.}
\begin{equation}\begin{split}
\amp_{04}(1^0,2^0,3^0,4^0) &= c_{04} \anb{12}\sqb{12}\anb{13}\sqb{13}\anb{23}\sqb{23}\\
\amp_{22}(1^+,2^-,3^0,4^0) &= c_{22} \anb{34}\sqb{34}\sql{1}3\anr{2}\sql{1}4\anr{2}\\
\amp_{40}(1^+,2^+,3^-,4^-)&= c_{40} \anb{12}\sqb{12}\anb{34}^2\sqb{12}^2.
\end{split}\end{equation}
The indices $n_\gamma$, $n_\varphi$ in $\amp_{n_\gamma n_\varphi}$ (or in the free couplings $c_{n_\gamma n_\varphi}$) denote the numbers of BI photons and sGals, respectively. For instance, the first amplitude $\amp_{04}$ corresponds to the scattering of zero BI photons and four sGals (it is thus a pure sGal amplitude).

\subsection{First iteration and 6-point amplitudes}
The first iteration means to construct the 6-point amplitudes by gluing the 4-point seed vertices and adding independent 6-point contact terms with unknown constants. The latter are to be determined using soft limits. The amplitude $\amp_{06}$ is a pure sGal amplitude, and it is well-known. For the 6-point amplitude $\amp_{24}$, we symbolically get
\begin{equation}
\amp_{24} = \amp_{22}\otimes_0\amp_{04} + \amp_{22}\otimes_1\amp_{22} + \sum_{i=1}^{\nu_{24}}c_{24,i}\amp_{24,i}^\mathrm{CT},
\end{equation}
where ``$\otimes_h$'' represents the gluing corresponding to an exchanged particle with the helicity\footnote{In our symbolic formulas, adding helicity-conjugated graphs (i.e. those with opposite helicity assignments to the external BI photon legs), which are present due to the helicity conservation, is implicit.} 
$h$ and the sum is over the $\nu_{24}=29$ independent 6-point contact counterterms. We have found that just demanding the $\bigo\rb{p^3}$ sGal soft limit for any scalar leg is enough to fix all the constants $c_{24,i}$ and also one of the two 4-point constants $c_{04}$ and $c_{22}$. The remaining unfixed 4-point constant represents the overall normalization of the amplitude.
Similarly, for the amplitude $\amp_{42}$, we get
\begin{equation}
\amp_{42} = \amp_{22}\otimes_0\amp_{22} + \amp_{22}\otimes_1\amp_{40} + \sum_{i=1}^{\nu_{42}}c_{42,i}\amp_{42,i}^\mathrm{CT},
\end{equation}
where $\nu_{42} = 42$. In this case, all $c_{42,i}$s and one of the two 4-point constants $c_{40}$ and $c_{22}$ can be fixed via imposing the multichiral soft limit \cite{Cheung:2018oki} for the BI photons\footnote{That means requiring $\amp_{42}=\bigo(t)$ when all the same helicity BI photons become soft, $p_i^\pm=\bigo(t)$ when $t\to 0$.} in addition to the $\bigo\rb{p^3}$ sGal limit.
Finally, for the amplitude $\amp_{60}$, we get $\nu_{60} = 5$ contact terms, and our procedure gives
\begin{equation}
\amp_{60} = \amp_{40}\otimes_1\amp_{40} + \sum_{i=1}^{\nu_{60}}c_{60,i}\amp_{60,i}^\mathrm{CT}.
\end{equation}
Since this amplitudes has no external scalars, we cannot use the sGal $\bigo\rb{p^3}$ limit. We have found that requiring $\amp_{60}=\bigo\rb{t^2}$ behavior in the limit where two BI photons with the same helicity become soft (i.e. a stronger requirement than in the previous case) is sufficient to fix all the $c_{60,i}$s.

\subsection{Higher iterations}
The $n$-point amplitudes for $n>6$ are iteratively constructed in a similar way, namely via gluing the lower-point contact vertices, adding a linear combination of independent $n$-point contact terms and then fixing as many as possible of the free constants by soft theorems. The resulting construction is schematically summarized in \rff{f:web}. Using this approach, we have proven numerically that $\amp_{08}$, $\amp_{26}$, and $\amp_{44}$ are uniquely fixed\footnote{Up to an overall normalization.} just by the sGal soft limit, and $\amp_{62}$ is not fixed by the sGal soft limit alone. Also, it appears that the amplitude $\amp_{80}$ cannot be uniquely fixed by any type of multichiral soft limit. This results seem to be in accord with the soft BCFW recursion. Indeed, assuming that the theory exists, the amplitudes $\amp_{n_\gamma n_\varphi}$ satisfying $n_\gamma < n_\varphi + 2$ could be recursively reconstructed using the soft BCFW recursion \cite{Cheung:2015ota} based on the sGal soft limit alone provided all the amplitudes with a smaller $n=n_\gamma+n_\varphi$ are already fixed,\footnote{Cf. also a similar discussion in \cite{Bonifacio:2019rpv}.} and the amplitudes $\amp_{n_\varphi+2,n_\varphi}$ could be reconstructed adding just one extra soft condition (e.g. some multichiral soft limit). 

\begin{figure}[ht]
\begin{center}
\begin{tikzpicture}[x=50pt, y=35pt, scale=.9]
\def\dgreen{green!50!black}
\node (04) at (-1,3) {$\stackrel{1}{\amp_{04}}$};
\node (22) at (0,3) {$\stackrel{1}{\amp_{22}}$};
\node (40) at (1,3) {$\stackrel{1}{\amp_{40}}$};

\node (06) [text=\dgreen] at (-1.5,2) {$\stackrel{5}{\amp_{06}}$};
\node (24) [text=\dgreen] at (-.5,2) {$\stackrel{29}{\amp_{24}}$};
\node (42) [text=orange] at (.5,2) {$\stackrel{42}{\amp_{42}}$};
\node (60) [text=red] at (1.5,2) {$\stackrel{5}{\amp_{60}}$};

\node (08) [text=\dgreen] at (-2,1) {$\stackrel{53}{\amp_{08}}$};
\node (26) [text=\dgreen] at (-1,1) {$\stackrel{696}{\amp_{26}}$};
\node (44) [text=\dgreen] at (0,1) {$\stackrel{2152}{\amp_{44}}$};
\node (62) [text=red] at (1,1) {$\stackrel{1280}{\amp_{62}}$};
\node (80) [text=red] at (2,1) {$\stackrel{94}{\amp_{80}}$};

\node (010) [text=\dgreen] at (-2.5,0) {$\stackrel{?}{\amp_{0,10}}$};
\node (28) [text=\dgreen] at (-1.5,0) {$\stackrel{?}{\amp_{28}}$};
\node (46) [text=\dgreen] at (-.5,0) {$\stackrel{?}{\amp_{46}}$};
\node (64) [text=orange] at (.5,0) {$\stackrel{?}{\amp_{64}}$};
\node (82) [text=red] at (1.5,0) {$\stackrel{?}{\amp_{82}}$};
\node (100) [text=red] at (2.5,0) {$\stackrel{?}{\amp_{10,0}}$};

\draw [->] (04) -- (06);
\draw [->] (04) -- (24);
\draw [->] (22) -- (24);
\draw [->] (22) -- (42);
\draw [->] (40) -- (42);
\draw [->] (40) -- (60);

\draw [->] (06) -- (08);
\draw [->] (06) -- (26);
\draw [->] (24) -- (26);
\draw [->] (24) -- (44);
\draw [->] (42) -- (44);
\draw [->] (42) -- (62);
\draw [->] (60) -- (62);
\draw [->] (60) -- (80);

\draw [->] (08) -- (010);
\draw [->] (08) -- (28);
\draw [->] (26) -- (28);
\draw [->] (26) -- (46);
\draw [->] (44) -- (46);
\draw [->] (44) -- (64);
\draw [->] (62) -- (64);
\draw [->] (62) -- (82);
\draw [->] (80) -- (82);
\draw [->] (80) -- (100);

\draw [->] (-1.6,3) -- (-3.1,0) node[midway,sloped,above] {Special Galileon};
\draw [->] (1.6,3) -- (3.1,0) node[midway,sloped,above] {Galileon-like BI};
\end{tikzpicture}
\end{center}
\vspace{-10pt}
\caption{The web of the amplitudes and the contact terms. We use the notation $\stackrel{\nu_{n_\gamma n_\varphi}}{\amp_{n_\gamma n_\varphi}}$, where $\amp_{n_\gamma n_\varphi}$ is either the set of contact term, or the amplitude, and $\nu_{n_\gamma n_\varphi}$ is the number of independent contact terms contributing to the amplitude $\amp_{n_\gamma n_\varphi}$. Whenever two nodes of the web can be connected by an oriented path, the contact terms corresponding to the starting point of the path contribute to the amplitude attached to the endpoint of the path. Provided the theory exists, the green-colored amplitudes can be uniquely reconstructed recursively from the sGal soft limit alone, the red-colored ones should be fixed by some additional requirement. The orange-colored amplitudes $\amp_{n_\varphi+2,n_\varphi}$ could be reconstructed using just one such extra soft limit.}
\label{f:web}
\end{figure}
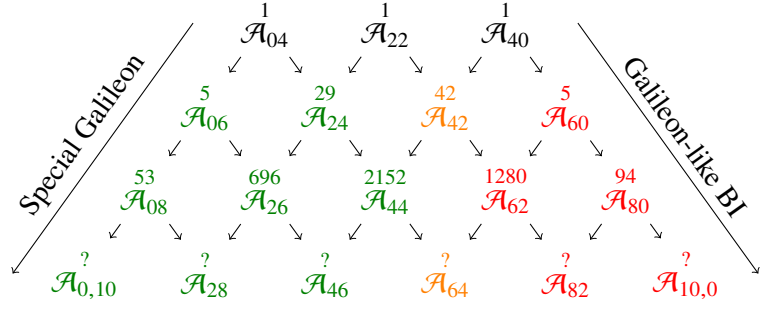

\section{Lagrangian approach}
On the Lagrangian level, the sufficient condition for the sGal soft limit is the invariance of the action with respect to the generalized polynomial shift symmetry\footnote{Here $\alpha$ is the free parameter of the sGal Lagrangian, see \cite{Novotny:2016jkh,Preucil:2019nxt} for details and for explicit formulas.} (here $G^{\alpha\beta}=G^{\beta\alpha}$, and $G^\mu_{\ \mu}=0$)
\begin{equation}
\delta\varphi = -\frac{1}{2}G^{\alpha\beta}\rb{\alpha^2 x_\alpha x_\beta + \prt_\alpha\varphi\prt_\beta\varphi}, 
\label{e:transf}
\end{equation}
The building blocks \cite{Novotny:2016jkh} are the effective metric $g_{\mu\nu}$, the extrinsic curvature $\ctns_{\alpha\mu\nu}$, and the scalar $\sigma$
\begin{equation}
g_{\mu\nu} = \eta_{\mu\nu} + \frac{1}{\alpha^2}\prt_\mu\prt\varphi\cdot\prt\prt_\nu\varphi,
\hspace{17pt}
\ctns_{\mu\nu\alpha} = -\frac{1}{\alpha}\prt_\mu\prt_\nu\prt_\alpha\varphi,
\hspace{17pt}
\sigma = \frac{1}{2\cu}\ln\frac{\det\rb{\eta+\frac{\cu}{\alpha}\prt\prt\varphi}}{\det\rb{\eta-\frac{\cu}{\alpha}\prt\prt\varphi}}.
\end{equation}
Any theory with even-point amplitudes only obeying the sGal soft limit and with the power counting $\varrho=2$ which couples the BI photon to the sGal has to include the following minimal Lagrangian
\begin{equation}
\mathcal{L}_\mathrm{min} = \mathcal{L}_\mathrm{sGal}[\varphi] - \frac{1}{4}\sqrt{|g|} V_0(\sigma) F_{\mu\alpha}F_{\nu\beta}g^{\mu\nu}g^{\alpha\beta},
\end{equation}
where $\mathcal{L}_\mathrm{sGal}[\varphi]$ is the sGal Lagrangian, and $V_0(\sigma)=V_0(-\sigma)$ is an arbitrary real function\footnote{See also \cite{Bonifacio:2019rpv}, where the special case $V_0=1$ has been studied.}. The action based on $\mathcal{L}_\mathrm{min}$ is invariant with respect to \rf{e:transf}, provided the BI photon field $A_\mu$ transforms as
\begin{equation}
\delta A_\mu = -G^{\alpha\beta}\rb{\prt_\alpha\varphi\prt_\beta A_\mu + A_\alpha\prt_\beta\prt_\mu\varphi}.
\end{equation}
However, $\mathcal{L}_\mathrm{min}$ alone gives a vanishing seed amplitude $\amp_{40}=0$, and the amplitude $\amp_{42}$ does not obey the multichiral soft limit. To reproduce the results of \rfs{s:bstr}, it is thus necessary to add also non-minimal invariants (starting from 4-photon terms) of the schematic form $V_1 F^4\ctns^2$ (12 independent terms), $V_2 (\cder\ctns) F^4$ (4 terms), $V_3 (\cder F)^2 F^2$ (9 terms), $V_4 (\cder F) F^3 \ctns$ (10 terms), and $V_5 (\cder^2 F) F^3$ (5 terms), where $\cder$ is the covariant derivative associated with the metric $g_{\mu\nu}$, and $V_i(\sigma)=(-1)^{i+1}V_i(-\sigma)$ are arbitrary functions. The resulting Lagrangian has infinitely many free couplings. Some combinations of them can be fixed by the multichiral soft limit applied to $\amp_{42}$.

\section{Conclusion}
We have presented the preliminary results of the application of the bootstrap method to the construction of unique tree amplitudes for a parity-and-helicity-conserving theory where the sGal is coupled to the BI photon. We have successfully fixed (up to one normalization constant) all the 6-point amplitudes by soft limits. Continuing to higher amplitudes, though some amplitudes are fixed, we have not found any appropriate constraints to make all the 8-point amplitudes unique. Also, a compatibility check of the first iteration with the second one has to be performed yet.

We also give an overview of a possible Lagrangian description of such a theory, based on covariant building blocks. Apart from the minimal (2-photon) part which must be always present, we have also performed a full classification of possible 4-photon non-minimal terms. The Lagrangian then reproduces the bootstrap results up to 6-points. It also allows for generalizations with non-vanishing odd-point amplitudes and violation of the helicity conservation.

\acknowledgments{This work was supported by the Czech Science Foundation, project No. GA18-17224S, by the Czech Ministry of Education, Youth and Sports, project No. LTAUSA17069, and by the Charles University Grant Agency, project No. 1108120.}

\bibliographystyle{JHEP}
\bibliography{proceedings.bib}
\end{document}